\documentclass[12pt,preprint]{aastex}
\usepackage{amssymb,natbib,epsfig}

\slugcomment{Accepted for publication in ApJ}

\shorttitle{Star-disk Interaction in NGC 2264 and Orion}
\shortauthors{Cieza \& Baliber}

\begin{document}


\title{Testing the Disk Regulation Paradigm with Spitzer Observations. II. \\
A Clear Signature of Star-Disk Interaction in NGC 2264 and the Orion Nebula Cluster}


\author{Lucas Cieza}
\affil{Astronomy Department, University of Texas at Austin \\ 
       1 University Station C1400, Austin, TX 78712}
\email{lcieza@astro.as.utexas.edu}

\and

\author{Nairn Baliber}
\affil{Las Cumbres Observatory Global Telescope and \\
       University of California, Santa Barbara\\
       6740 Cortona Dr. Ste 102, Goleta, CA 93117}
\email{baliber@lcogt.net}

\begin{abstract}
Observations of PMS star rotation periods reveal slow rotators in
young clusters of various ages, indicating that angular momentum is
somehow removed from these rotating masses.  The mechanism by which
spin-up is regulated as young stars contract has been one of the
longest-standing problems in star formation.  Attempts to
observationally confirm the prevailing theory that magnetic
interaction between the star and its circumstellar disk regulates
these rotation periods have produced mixed results.  In this paper, we
use the unprecedented disk identification capability of the {\it
Spitzer} Space Telescope to test the star-disk interaction paradigm in
two young clusters, NGC 2264 and the Orion Nebula Cluster (ONC). We
show that once mass effects and sensitivity biases are removed, a
clear increase in the disk fraction with period can be observed in
both clusters across the entire period range populated by cluster
members.  We also show that the long-period peak (P $\sim$8 days) of
the bimodal distribution observed for high-mass stars in the ONC is
dominated by a population of stars possessing a disk, while the
short-period peak (P $\sim$2 days) is dominated by a population of
stars without a disk.  Our results represent the strongest evidence to
date that star-disk interaction regulates the angular momentum of
these young stars.  This study will make possible quantitative
comparisons between the observed period distributions of stars with
and without a disk and numerical models of the angular momentum
evolution of young stars.

\end{abstract}


\section{Introduction}
For many years, the loss of angular momentum in the evolution of
pre-main-sequence (PMS) stars was a fundamental problem in the theory of star
formation.  As PMS stars contract by a factor of $\sim$\,2\,--\,3 during their
first 3\,Myrs of evolution, models assuming homologous contraction and
conservation of angular momentum dictate that all stars less than
$\sim$1.2\,${\rm M}_{\odot}$ should rotate with periods shorter than $\sim$2
days by an age of 2\,Myrs \citep{dantona98,herbst00a}.  However, observations
of clusters determined to be $\sim$2\,Myrs old or older show that most PMS
stars rotate much slower than expected.  Interaction between the magnetic
field of a young star and the inner regions of its protoplanetary disk has
been invoked by virtually every rotational evolution model as the mechanism by
which the rotation periods of these stars are regulated as they evolve onto
the main sequence \citep{konigl91,shu94,hartmann02,matt05}.

A first-order prediction of these models is stars interacting with their disks
should have longer rotation periods than stars that have already lost their
disks, leaving them free to spin up as they contract \citep{herbst05}.  Early
observations of rotation periods of PMS stars, obtained by monitoring the
brightness modulation produced by stellar surface features, seemed to support
the star-disk interaction scenario for angular momentum regulation.  

Some studies showed correlations between rotation period and ground-based disk
indicators \citep[most commonly excess K-band emission, e.g.][]{herbst02} and
interpreted these results as strong evidence for star-disk interaction,
claiming that stars with longer periods were rotating more slowly because they
had transferred angular momentum to their disks
\citep{edwards93,herbst00a,herbst02,lamm05}.  This interpretation suffers from
several problems as various issues can mask or mimic the correlation, such as
sample size, sensitivity biases, mass effects, and, most importantly,
ambiguous disk indicators.  Therefore, the correlation between infrared (IR)
excess and rotation period was challenged by studies which failed to find any
correlation between rotation period and a range of disk and accretion
indicators in various clusters
\citep{stassun99,rebull01,rebull04,makidon04,littlefair05}.  
The current work presents conclusive evidence that star-disk interaction
is the mechanism by which PMS angular momentum is regulated.

The ability to overcome the confusion surrounding the angular momentum problem
arrived with the acquisition of {\it Spitzer} mid-IR observations sensitive
enough to unambiguously determine the presence of a disk around PMS stars in
the IRAC band passes.  In a study of the young cluster IC\,348
\citep[][hereafter, Paper I]{cieza06}, we find that 8.0\,$\mu$m data are needed
to clearly identify all of the disks in a given sample of stars (fig. 5)
\citep[see also][]{hartmann05,rebull06}.  

Both the near infrared (NIR) and IRAC wavelengths trace the inner rim
of the disk \citep{allen04,cieza05}. However, since the
magnitude of the IR excess over the stellar photosphere is much larger
at IRAC wavelengths than it is at NIR wavelengths (e.g., 2MASS
wavelengths), IRAC observations, unlike ground-based studies, are
sensitive enough to unambiguously determine the presence of a disk
around PMS stars.

It might be argued that the sensitivity of the 8.0 $\mu$m excess
allows detection of both active, accreting disks and passive disks
which no longer interact with their parent stars.  However, the fact
that the presence of an IRAC excess does not guarantee that the stars
are {\it currently} accreting does not affect our conclusions for
multiple reasons.  First, disk dissipation timescales (the time it
takes for a given disk to dissipate once it has begun that process)
are much shorter than the mean accretion disk lifetimes.  For
instance, from the ratio of the number of weak-line T Tauri stars
(WTTS) with IRAC excesses to the number of PMS stars (WTTS plus
classical T Tauri Stars, CTTS), \citet{cieza07} estimate the
transition timescale from optically thick accretion disks to disks
undetectable at IRAC wavelengths to be 0.4 Myrs, a factor of $\sim$10
smaller than typical accretion disk lifetimes.  This timescale is very
similar to the transition timescale between an optically thick disk
and an optically thin disk found by \citet{skrutskie90} and
\citet{wolk96}, based on the number of ``transition objects,'' which
they define as targets without K-band excess but with strong IRAS
excesses.  Thus, if an object presently shows an IRAC excess, it is
much more likely than not to be accreting.

Moreover, once star-disk interaction stops, it takes some time for
stars to spin up significantly.  This ``reaction lag-time'' will tend
to compensate for the ``detectability lag-time'' (the time between the
end of star-disk interaction and the point at which the disk is no
longer detectable at IRAC wavelengths).  As a result, IRAC 8 $\mu$m
excess is arguably the best current method with which to reliably
detect the presence of an inner circumstellar disk and study the
effects of star-disk interaction on the angular momentum history of
young stars.

In this paper, we present a new study of the angular momentum history of two
young stellar clusters, the ONC ($\sim$1 Myr) and NGC 2264 ($\sim$2-3
Myrs). In \S\ref{previous}, we discuss results from previous work on these two
clusters and IC 348, the other well-studied PMS star cluster, detailing the
important factors, such as accurate disk identification and sample selection,
needed to isolate the effects of circumstellar disks on the current rotation
period distributions of young clusters.  In \S\ref{results}, we describe our
new results from NGC 2264, using rotation periods from the literature and
public {\it Spitzer} data, and a reanalysis of the data in the ONC study
presented by \citet{rebull06}, using the same mass sample for each study, with
a stricter sample selection in the ONC because of the different extinction
levels in that cluster. In \S\ref{discussion}, we compare the two results and
discuss robust Monte Carlo simulations, already underway, which will determine
what star and disk evolution parameters are allowed given the current rotation
period distributions of stars with and without a disk in each cluster.  Future
work on the subject is also proposed.  Our conclusions are summarized in
\S\ref{conclusion}.

\section{Previous Observational Results}\label{previous}
Three clusters have been studied extensively to produce rotation periods from
photometric monitoring campaigns, spectral types from spectroscopy and
photometric colors, and disk identification from various disk
indicators.  Until {\it Spitzer} mid-IR data became available, disk
identification was limited to ground-based color excesses, mostly in the
NIR, and H$\alpha$ equivalent widths.

These indicators, however, have not provided disk identification accurate
and reliable enough to study the effect circumstellar disks have had
throughout the lives of PMS stars at the age of these clusters.  The
correlation between rotation period and NIR color excess can be masked by
biases introduced by these data.  For instance, it has been shown that the
NIR disk indicator misses 30\% of the disks that can be detected at longer
wavelengths \citep{hillenbrand98}. Also, the ground-based photometry used to
calculate the excess most often comes from different epochs, which can affect
the results due to the high photometric variability of these stars
\citep{rebull01}.

Moreover, correlations between rotation period and
NIR excess can be caused by a secondary effect of the correlation between
mass and rotation period, in a sense mimicking the result expected from
star-disk interaction \citep{littlefair05}.  
As shown by previous work \citep{herbst00a,herbst02,lamm05,cieza06}, rotation
period distributions of PMS stars are highly dependent on mass.  PMS stars of
later spectral types behave differently than do stars M2 and earlier,
corresponding to masses of M $\ge$ 0.25M$_{\odot}$ at the ages of these
clusters, according to theoretical evolutionary tracks \citep{dantona98}
(hereafter, high-mass stars). 
ONC high-mass stars show a bimodal period distribution, while low-mass stars
rotate more rapidly than the high-mass stars, with a unimodal distribution
peaking at $\sim$2 days and a tail of longer periods \citep{herbst01}.  In NGC
2264, the high-mass stars do not show a clear bimodal distribution, but
they do rotate more slowly than the low-mass population.  The low-mass period
distribution is also more sharply peaked than the high-mass distribution.  Due
to the difference in the color contrast between the stellar photosphere
and the inner disk, NIR excess tends to be greater for high-mass stars
than for low-mass stars \citep{hillenbrand98}, and since lower mass stars tend
to rotate faster than higher mass stars, this can result in a correlation
between NIR excess and rotation period that is not necessarily connected
to star-disk interaction.  Therefore, these two different populations must be
studied separately.

{\it Spitzer}'s Infrared Array Camera (IRAC, 3.6--8.0 $\mu$m) \citep{fazio04}
allows the first observations sensitive enough at mid-IR wavelengths to
accurately determine the presence of inner circumstellar disks for a
statistically significant number of PMS stars with known rotation periods.
The observations at every IRAC wavelength are also observed concurrently,
overcoming previous limitations caused by stellar magnitude variations between
observations.  However, even with an accurate identification of circumstellar
disks, biases and selection effects can still mask useful measurements of the
effects of star-disk interaction on young stellar populations.  As discussed
in \S\ref{data} and \S\ref{results}, selecting a uniform and complete sample
of stars is critical in order to be able to detect the role star-disk
interaction has on the angular momentum evolution of PMS stars.  What follows
is a more detailed discussion of the mixed results of
previous searches for the effects of star-disk interaction on rotation
period distributions in each cluster.

\subsection{NGC 2264}
 
Virtually all $\sim$500 known rotation periods in NGC 2264 can be
collected from two studies \citep{makidon04,lamm05}.
\citet{makidon04} conduct a study of 201 stars in this cluster. They
examine the hypothesis that star-disk interaction regulates the
angular momentum of PMS stars by searching for a correlation between
rotation period and 4 different disk indicators (U-V, I$_C$--K$_s$,
H--K$_s$, and H$\alpha$). They find ``no conclusive evidence that more
slowly rotating stars have disk indicators, or that faster rotating
stars are less likely to have disk indicators.''  \citet{lamm05}
investigate the disk regulation hypothesis in NGC 2264 by studying
rotation period distributions of CTTS and WTTS using a
R$_C$--H$\alpha$ vs. R$_C$--I$_C$ color criterion to distinguish
between the two populations.  They find that the distribution of
rotation periods of the high-mass CTTS and WTTS populations ``looks
quite different'' even though ``the statistics are poor.''  Namely,
according a Kolmogorov-Smirnov test, they find that there is a 0.02
probability that the distribution of high mass CTTS and WTTS are
equivalent.  They attribute this marginally significant difference in
the rotation period distribution to the fact that the stars classified
as CTTS are more likely to have a disk than those classified as WTTS.
Although they attempt to separate stars by both mass and disk
presence, their results are hampered by an inefficient disk
identifier.  Fewer than half of their high-mass stars, for example,
are used in their analysis (72 out of 184), leaving 112 stars as
ambiguous disk identifications.

\subsection{ONC}
With over 900 known rotation periods, the ONC has been the focus of most
studies of the angular momentum evolution of PMS stars.  \citet{edwards93}
conduct a study of 34 T Tauri stars from Taurus and the ONC, ranging in
spectral type from K7 to M1 and in age from 1 to 10 Myrs. They show a
correlation between H-K excess and rotation period for these stars; however,
in a subsequent study by \citet{stassun99} for stars in the ONC, no such
correlation was found.  \citet{herbst00a} find a strong period dependence on
stellar mass and confirm the bimodal distribution for high-mass stars
originally suggested by an earlier study from that group \citep{attridge92}.
They explain that the previous study by \citet{stassun99} did not exhibit
the bimodal period distribution because its sample was dominated by low-mass
stars.  They find a ``weak but significant'' correlation with period among
stars with M $>$ 0.25 M$_\odot$, but argue that the strongest evidence for
disk-locking is the bimodal distribution itself.

\citet{rebull01} study 4 fields in the outer ONC, surrounding but not
including the Trapezium region. They conclude that ``There is no unambiguous
correlation of period with I$_C$-Ks, H-Ks, and U-V color excesses or more
indirect disk indicators; the slowest rotators are not necessarily the disk
candidates, and the disk candidates are not necessarily the slow rotators,
regardless of how one defines a disk candidate'' \citep{rebull01}.

Subsequently, \citet{herbst02} show a correlation between rotation period and
I-K excess for stars in the ONC which they claim has a ``very high'' level of
significance. They find that slower rotators with periods $>$6.28 d show a
mean IR excess emission, $\Delta$(I--K) = 0.55 $\pm$0.05, 
and more rapidly rotating stars
with periods $<$3.14 d have a mean $\Delta$(I--K) of 0.17 $\pm$0.05.  With
this result, they claim, ``The long-suspected, but somewhat controversial,
correlation between rotation and excess infrared emission, which is relevant
to the disk-locking hypothesis, is finally confirmed at a very high
significance level. There is no doubt now that more slowly rotating stars in
the ONC have, on average, greater infrared excess emission than do their more
rapidly rotating counterparts'' \citep{herbst02}.  This correlation between
rotation and the magnitude of the excess infrared emission has been
interpreted by the Herbst group as conclusive support of the star-disk
interaction hypothesis \citep{herbst02,herbst06}.

However, having failed to find a correlation between period and disk
indicators such as H$\alpha$ emission, U-V color excess, and K-L color excess,
other groups have a different interpretation of the \citet{herbst02} results
and argue that the observed correlation between period and the magnitude of
the infrared excess does not represent strong support for star-disk
interaction.  \citet{makidon04} state that ``the size (and indeed the
presence) of the NIR excess need not be well correlated with the presence
of a circumstellar disk owing to the combined effects of inclination and inner
disk hole effects (see Hillenbrand et al. 1998; Mathieu 2003). Therefore,
correlations between period and NIR excess strength are not necessarily
particularly meaningful'' \citep{makidon04}.  \citet{littlefair05} express
similar concerns about the \citet{herbst02} result, and add an additional one:
a mass effect. They argue that the ``the level of I--K excess depends upon a
number of factors: disc mass, inclination angle, inner disc hole size and disc
structure. It also depends strongly upon stellar mass, with excesses around
high-mass stars being much stronger than excesses around low-mass stars
(Hillenbrand et al. 1998).  What this means for the ONC is that the low-mass
stars, which are rotating rapidly, will necessarily exhibit smaller I--K
excesses than the high-mass stars, which rotate more slowly. Such an effect
could easily be responsible for the apparent correlation between rotation rate
and infrared excess. The claims that the ONC offers strong support for disc
locking should therefore be interpreted with caution'' \citep{littlefair05}.

We note that the slow and fast rotators ($\omega$ $<$ 1 radian/d and $\omega$
$>$ 2 radian/d, corresponding to periods longer than 6.28 days and shorter
than 3.14 days, respectively), for which the \citet{herbst02} study finds very
different IR-excess distributions, indeed have very different mass
distributions. In their short-period sample, there are twice as many low-mass
stars as high-mass stars (80 vs. 40), while in their long-period sample, there
are only 34 low-mass stars vs. 84 high-mass stars.  Moreover, the mean IR
excess emission for long-period stars in the ONC found by the Herbst et
al. study ($\Delta$(I--K) = 0.55) is increased significantly by the 8 stars
(all high-mass) with the highest excesses in the group.
                                                                           
Recently, using {\it Spitzer} IRAC data as a more reliable disk indicator,
\citet{rebull06} studied the angular momentum problem in the ONC central and
surrounding regions using periods from the literature.  Having been able to
accurately determine which stars have disks and which do not, they were able
to show a connection between stellar rotation and the presence of a
circumstellar disk. In particular, they show that stars with periods shorter
than 1.8 days are significantly less likely to have a disk that stars with
periods longer than 1.8 days.  However, they also find that ``among the slower
rotators (stars with periods $>$ 1.8 days), the period distributions for stars
with and without disks ([3.6]--[8] $>$ 1 and $<$ 1, where bracketed notation
indicates IRAC colors) are statistically indistinguishable.'' Though
suggestive, by itself this result does not lend conclusive support to the
star-disk interaction scenario because the short-period objects that represent
the correlation make up less than 20\% of the entire population, leaving over
80\% of the objects showing no correlation.  Also, this 1.8-day period cut is
arbitrary, chosen in order to maximize the result rather than for a specific
scientific reason.  Other factors, such as an overabundance of close binaries
among fast rotators, could also account for the low disk fraction in this
population \citep{rebull06,cieza06}.  As shown in \S\ref{data} and
\S\ref{results}, even when using {\it Spitzer} colors as a disk indicator, a
general correlation between disk fraction and rotation period can only be seen
across the entire period range after the mass effects and sensitivity biases
are removed from the sample.

\subsection{IC 348}

To date, only two groups have searched for a period-disk correlation in IC
348 ($\sim$3\,Myrs).  \citet{littlefair05} study a sample of 50 periodic stars
and search for a correlation between period and K-L color excess (available
for 30 stars) and H$\alpha$ (available for 43 stars), but find no significant
correlation.

Thanks to a very deep IRAC GTO survey (1600 sec exposures/pixel),
IC\,348 is the only cluster of the three discussed in this paper that
currently has deep-enough observations to reach the photospheric level
of the entire sample of periodic stars at all four IRAC
wavelengths. In Paper I, a study of IC 348 using these {\it Spitzer}
data, we find a similar result to the one reported by \citet{rebull06}
in the ONC, although with a smaller level of significance given the
size of our sample.

Namely, in a total sample of $\sim$130 stars with rotation periods, we find a
small subset of cluster members that rotate with periods shorter than $\sim$2
days, showing a significantly lower disk fraction than the rest of the cluster
population.  We also find no statistically significant difference between the
rotation period distribution of stars with and without disks at periods longer
than $\sim$2 days. We analyze the populations of stars with and without disks
regardless of stellar mass and the populations of high-mass and low-mass stars
independently.  When the entire sample is considered, we find a bimodal
distribution of periods for the stars with disks which offers no support for
star-disk interaction.  However, after subdividing by mass the population of
stars with and without a disk, there are too few stars in each mass
bin for the disk and no-disk population to make the analysis as a function of
period statistically meaningful (see \S\ref{meffect}).

\section{The Data}\label{data}
\subsection{NGC 2264}

\subsubsection{NGC 2264 Rotation Periods}
\citet{makidon04} report rotation periods for 201 stars. Based on
their false alarm probability levels, they divide their periods into two
quality categories, 1 and 2.  \citet{lamm05} report rotation periods for 405
stars. We combined the 114 ``quality 1'' rotation periods reported by
\citet{makidon04} and the 405 rotation periods from \citet{lamm05}.  There
were 74 stars in common between these two groups of 114 and 405 stars, which
means that we have selected a total of 445 individual stellar rotation
periods.  We list the periods of these stars in Table 1, along
with their coordinates and the R$_C$ and I$_C$ photometry reported by the two
groups.  We adopt the periods from the \citet{lamm05} study for the 74 stars
common to both studies.  Their work shows that $\sim$95$\%$ of these 74
stellar rotation periods are identical to those of the ``quality 1'' periods
reported by the Makidon group, highlighting the reliability of all of the
periods listed in Table 1.

\subsubsection{NGC 2264 {\it Spitzer} data}
NGC 2264 was observed with IRAC \citep{fazio04} as part of
the {\it Spitzer} Guaranteed Time Observation program 
``Disk Evolution in the Planet Formation Epoch'' (PID=37).  The observations
consist of 4 dithers and were conducted in the High Dynamic Range mode
which includes 0.4 sec observations before 10.4 sec exposures at each
dither position. This mode allows photometry of both bright and faint
stars at the same time.  Each dither consists of 7$\times$11 IRAC
fields with 290$''$ offsets, resulting in a total mapped area of
$\sim$33$'$$\times$51$'$ at each of the IRAC wavelengths (3.6, 4.5,
5.8, and 8.0 $\mu$m).  See \citet{young06} for a more detailed description
of the IRAC observations of NGC 2264.  We retrieved from the {\it Spitzer}
Science Center (SSC) archive the Basic Calibrated data of NGC 2264
that was processed with the SSC pipeline version S11.0.2. The
Astronomical Observation Request (AOR) Keys of the data are
0003956480, 0003956992, 0003956736, and 0003957248. We mosaicked the
IRAC data and produced point-source catalogs for each band using the
pipeline developed as part of the {\it Spitzer} Legacy Project, ``From
Molecular Cores to Planet-forming Disks'' (c2d).  See
\citet{evans06} for a detailed description of the c2d
pipeline.

\subsubsection{NGC 2264 Sample Selection}
Obtaining an unbiased sample is critical in order to be able to study
the connection between circumstellar disks and PMS star angular
momentum evolution.  Using {\it Spitzer} data in our study of IC 348
(Paper I), we show that $\sim$\,40$\%$ of the circumstellar disks
identified with 5.8 and/or 8.0\,$\mu$m excesses in IC\,348 show no
clear excess at shorter wavelengths, indicating the necessity of 8.0
$\mu$m data for every star in the sample to prevent missing disks
which cannot be detected at shorter wavelenghts.

We searched our point source catalogs for IRAC fluxes of the periodic
stars in NGC 2264 listed in Table 1.  We found 3.6 $\mu$m fluxes for
all 445 of the objects, and data for 436, 371, and 229 stars at 4.5,
5.8, and 8.0 $\mu$m respectively.  The fact that both \emph{Spitzer's}
sensitivity and photospheric fluxes decrease with increasing
wavelength explains the smaller number of detections at 5.8 and 8.0
$\mu$m.  Following Paper I, we use the [3.6]-[8.0] colors for disk
identification purposes ([3.6]-[8.0] $<$ 0.7 represents a bare stellar
photosphere, and [3.6]-[8.0] $>$ 0.7 a star with a disk); therefore,
in order to preserve the reliability of our disk identification, we
restrict our analysis to the 229 stars with available 3.6 and 8.0
$\mu$m fluxes.

\subsubsection{Mass Bias}

\begin{figure}
\epsscale{1.0}
\plottwo{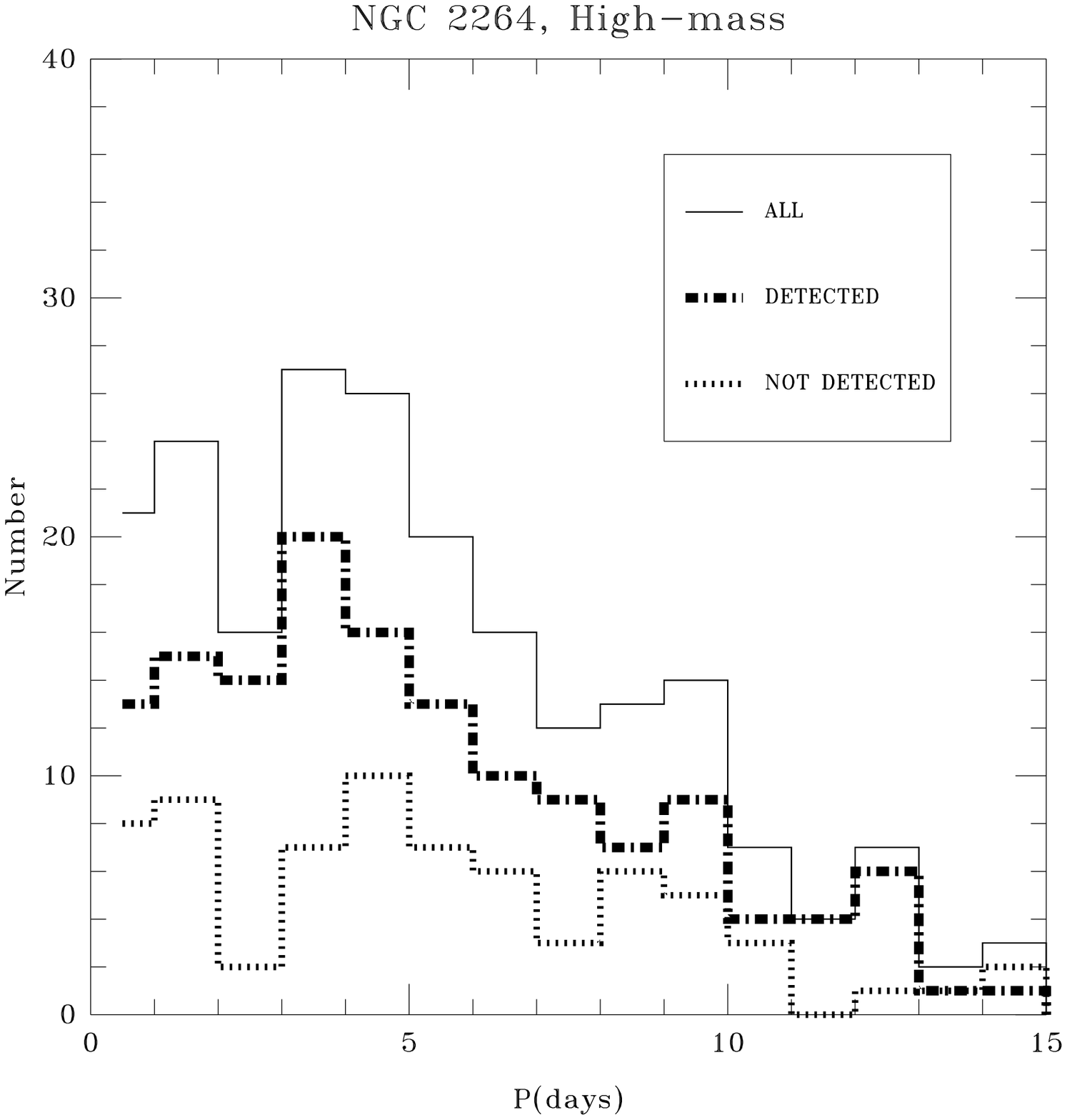}{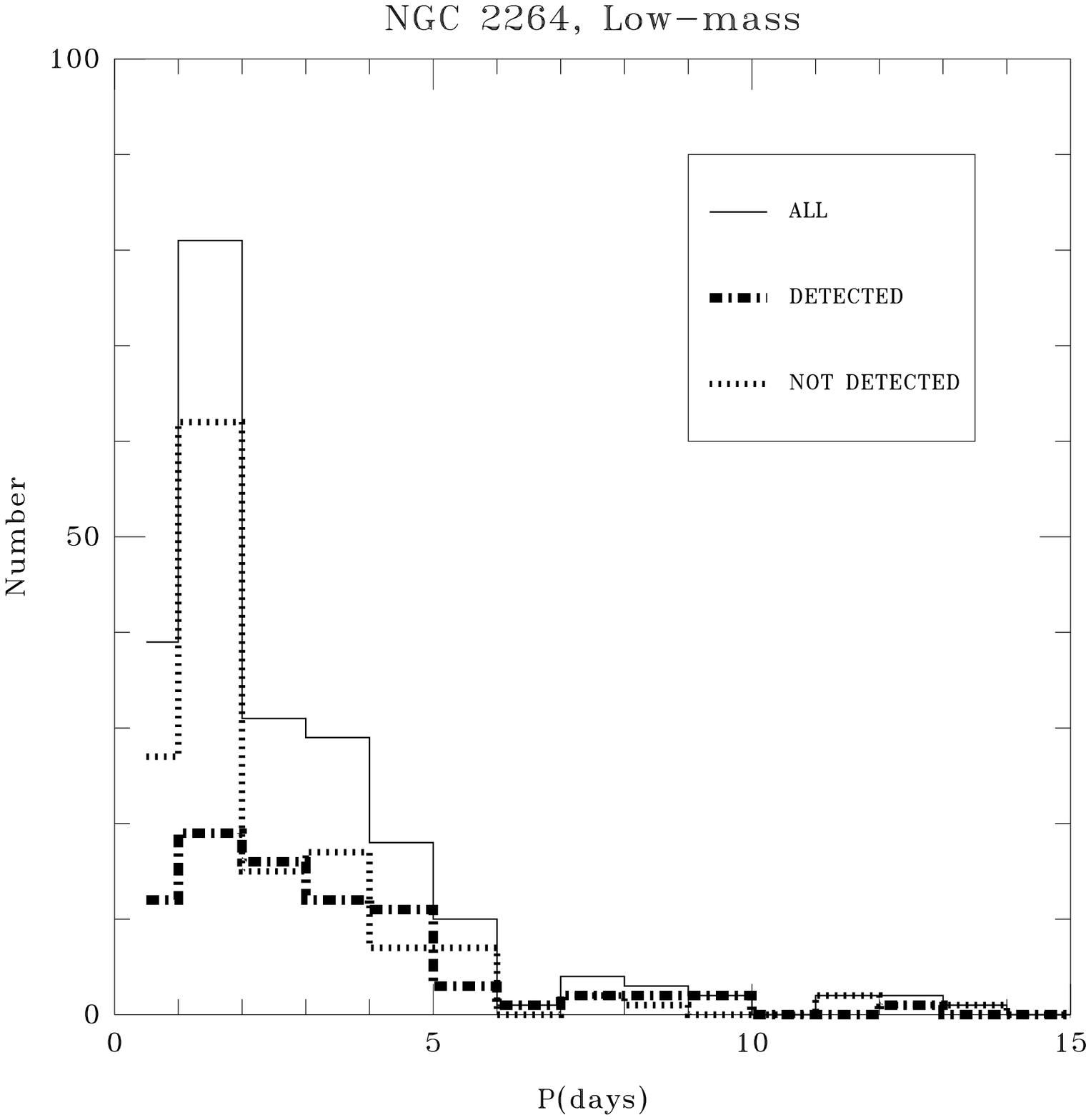}
\caption{{\bf Mass-segregated period histograms for stars with and without 8.0
    $\mu$m data in NGC 2264.} {\bf (Left Panel)} Period histogram for
    high-mass stars ([R-I] $<$ 1.3) in NGC 2264. The three different lines
    represent all stars (n=212, solid line), stars detected with {\it
    Spitzer's} IRAC instrument at 8.0 $\mu$m (n=142, dot-dash line), and stars
    not detected at 8.0 $\mu$m (n=70, dotted line). {\bf (Right Panel)} Period
    histogram for low-mass stars ([R-I] $>$ 1.3) in the cluster. The three
    different lines represent all stars (n=223, solid line), stars detected
    with {\it Spitzer's} IRAC instrument at 8.0 $\mu$m (n=81, dot-dash line),
    and stars not detected at 8.0 $\mu$m (n=142, dotted line).  As previously
    noted by \citet{lamm05} in NGC 2264 and \citet{herbst02} in the core of
    the ONC, low- and high-mass stars have clearly different period
    distributions.  Since the 8.0 $\mu$m data is needed for a reliable disk
    identification \citep{rebull06,cieza06}, our analysis is restricted to
    stars detected at this wavelength. The period distribution of the
    high-mass stars detected at 8.0 $\mu$m is statistically indistinguishable
    from that of the undetected stars (P = 0.96, Kolmogorov-Smirnov two-sample
    test, n1 = 142, n2 = 70).  In contrast, the period distribution of
    low-mass stars detected at 8.0 $\mu$m is significantly different than that
    of the undetected stars (P = 1.6e-3, K-S two-sample test, n1 = 81, n2 =
    142).  As the low-mass sample has a much lower detected fraction of stars
    at shorter periods than at longer periods, the biases in this sample
    prevent us from using it.\label{NGCdetect}}
\end{figure}

Because IRAC 8.0 $\mu$m data are required for reliable disk identification, a
mass bias can be introduced due to the sensitivity limits of those data in a
magnitude-limited sample.  To illustrate this effect, in Fig.~\ref{NGCdetect}
we plot histograms of the period distributions of the high- and low-mass stars
in the NGC 2264 data set that were detected at 8.0 $\mu$m and those that were
not.  In the left panel, the period distribution of the high-mass stars
detected at 8.0 $\mu$m is statistically indistinguishable from the
distribution of stars with no 8.0 $\mu$m detection (P = 0.96,
Kolmogorov-Smirnov two-sample test, n1 = 142, n2 = 70).  In the right panel,
on the other hand, the period distribution of low-mass stars detected at 8.0
$\mu$m is significantly different than that of the undetected stars (P =
1.6e-3, K-S two-sample test, n1 = 81, n2 = 142).  The fraction of stars with
periods $<$2 days detected at 8.0 $\mu$m is 26\%, and the fraction with P $>$
2 days detected is 46\%.  In the low-mass sample, there is a strong bias
against the fastest rotators, which is expected if these stars are faint
because they are the lowest-mass stars in the sample and/or they
preferentially have no disks.  Also, if optical colors are used to estimate
mass, the low-mass sample can be contaminated by highly extincted high-mass
stars, which have a different rotation period distribution than lower-mass
stars.

Since low-mass stars suffer from these two effects which render the
current sample unreliable, we must segregate the sample by spectral
type (corresponding to masses given by evolutionary tracks).  Since
brighter high-mass stars do not suffer as much from these sources of
sample bias and contamination (reddened low-mass stars cannot be
mistaken for high-mass stars), in addition to requiring {\it Spitzer}
8.0 $\micron$ detection, we further restrict our study to the
high-mass population.

Spectral types to estimate masses are available for only a fraction of
the stars in NGC 2264.  However, NGC 2264 has relatively low
extinction (A$_{V}$ $\sim$0.5 mag) \citep{rebull01}, which is fairly
uniform across the field of view covered by this study.  This allows
us to use R-I colors to make a mass cut and retain a relatively
uncontaminated sample of high-mass stars.  We use an R-I color $<$ 1.3
(corresponding to unextincted M2 stars and stars with earlier spectral
types \citep{kenyon95}) as the cutoff for this sample.  Restricting
our sample, as discussed, based on {\it Spitzer} data and stellar
mass, leaves a final sample of 142 high-mass stars with known rotation
periods detected by IRAC at 8.0 $\micron$ with which we search for a
correlation between rotation period and the presence of a
circumstellar disk.

\subsection{The ONC}
For our analysis of the Orion Nebula Cluster and its surroundings, we
combine in Table 2 the rotation periods and {\it Spitzer} data
presented by \citet{rebull06} with spectral types from the literature
\citep{rebull01,hillenbrand97}. The study by \citet{rebull06} covers
the intersection of the IRAC maps of the Orion star-forming complex
(total area $\sim$ 6.8 deg$^2$) and the Orion regions containing stars
with known rotation periods.  These regions are the ONC (i.e. the
region within the $\sim$20$'$ of the Trapezium) and the ``Flanking
Fields'' (four 45$'$ $\times$ 45$'$ fields centered $\sim$30$'$ East,
West, North, and South of the ONC). The study of the ONC by
\citet{hillenbrand97} provides spectral types for $\sim$70$\%$ of the
periodic stars studied by \citet{rebull06} that are located within
their 34$'$ $\times$ 36$'$ ONC field.  \citet{rebull01} provides
spectral types for only $\sim$30$\%$ of the periodic stars studied by
\citet{rebull06} that are located in the Flanking Fields.

\subsubsection{ONC Sample Selection}
A large fraction of the stars in the \citet{rebull06} sample does not have
measured spectral types.  For stars with no spectral type measurement,
\citet{rebull06} make a mass cut by placing these stars on I vs. (V-I)
color-magnitude diagrams.  Unlike in the case of NGC 2264, this method is
unreliable for the ONC sample because the extinction is high and highly
variable (A$V$$\sim1-5$) across the entire field of view covered by the
study \citep{hillenbrand97}.  Using colors for spectral type classification can
lead to a blending of the period distribution of the high- and low-mass stars
(see \S\ref{meffect} for a more detailed discussion).
We therefore limit our analysis of the \citet{rebull06} sample of stars with
known rotation periods to those with measured spectral types (M2 and earlier).
This leaves 133 high-mass stars with which to monitor the effects of star-disk
interaction.

\section{Results}\label{results}
\subsection{NGC 2264}
\begin{figure}
\epsscale{1}
\plotone{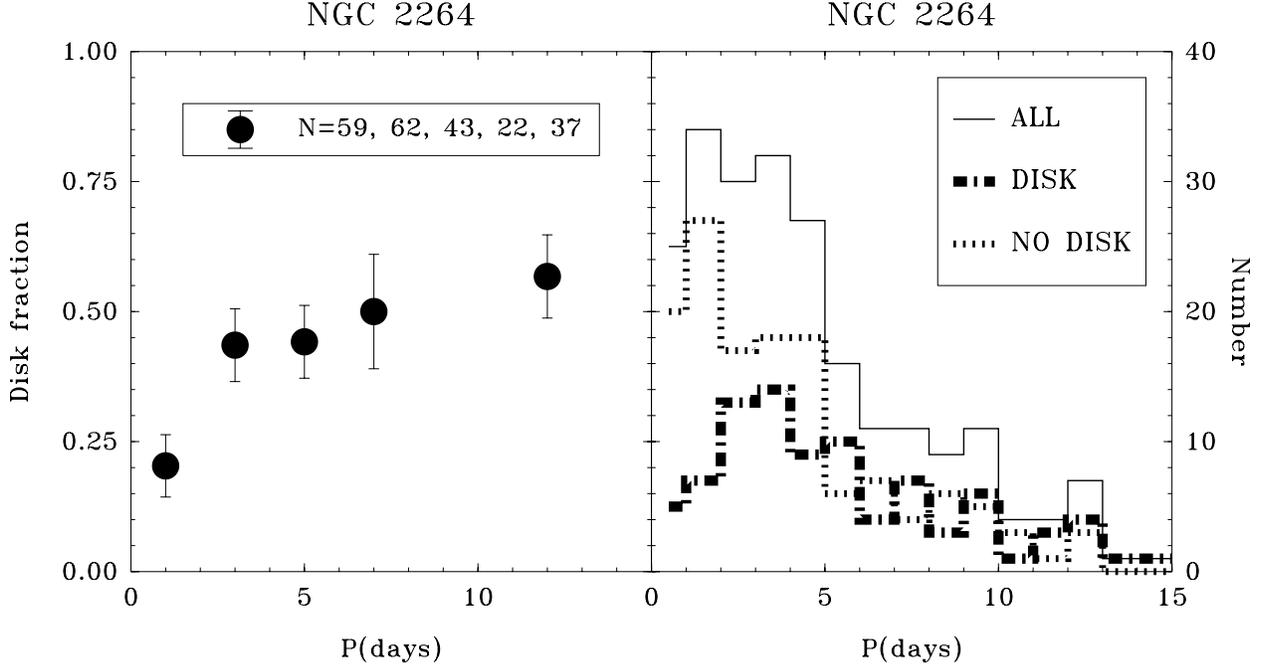}
\caption{{\bf Results for all stars in NGC 2264.} {\bf (Left Panel)} The disk
fraction as a function of period for the stars in NGC 2264 with rotation
periods $<$ 15 days and both 3.6 and 8.0 $\mu$m IRAC data, enough for an
accurate disk identification.  The error bars represent the 68\% confidence
level (1$\sigma$) of the measurements.  The only significant feature is the
lower disk fraction of the stars with shortest periods (P $<$ 2 days) with
respect to that of the rest of the sample.  {\bf (Right Panel)} The period
histogram for for the same sample of stars.  The three different lines
represent all stars (solid line), stars with IR-excess indicating the presence
of a disk (dot-dash line) and stars with no detected disk signature (dotted
line).  For periods longer than 2 days, the distribution of periods for stars
with and without a disk are statistically indistinguishable (P=0.211,
Kolmogorov-Smirnov two sample test, n1=76, n2=88).\label{NGCall}}
\end{figure}

When disk fraction is plotted as a function of period for all NGC 2264 members
without separating the populations by spectral type (Fig.~\ref{NGCall}, left
panel), we find that the only significant feature is a lower disk fraction
(17$\pm$4$\%$) for stars in the shortest period bin (P $\leq$ 2 days) compared
to that of the rest of the sample (45$\pm$4$\%$).  For periods longer than 2
days, the distributions of periods for stars with and without a disk
(Fig.~\ref{NGCall}, right panel) are statistically indistinguishable (P=0.211,
Kolmogorov-Smirnov two sample test, n1=76, n2=88). This is, in essence, an
identical result to those found in the {\it Spitzer} studies of the ONC
\citep{rebull06} and IC 348 \citep{cieza06}.

\begin{figure}
\epsscale{1}
\plotone{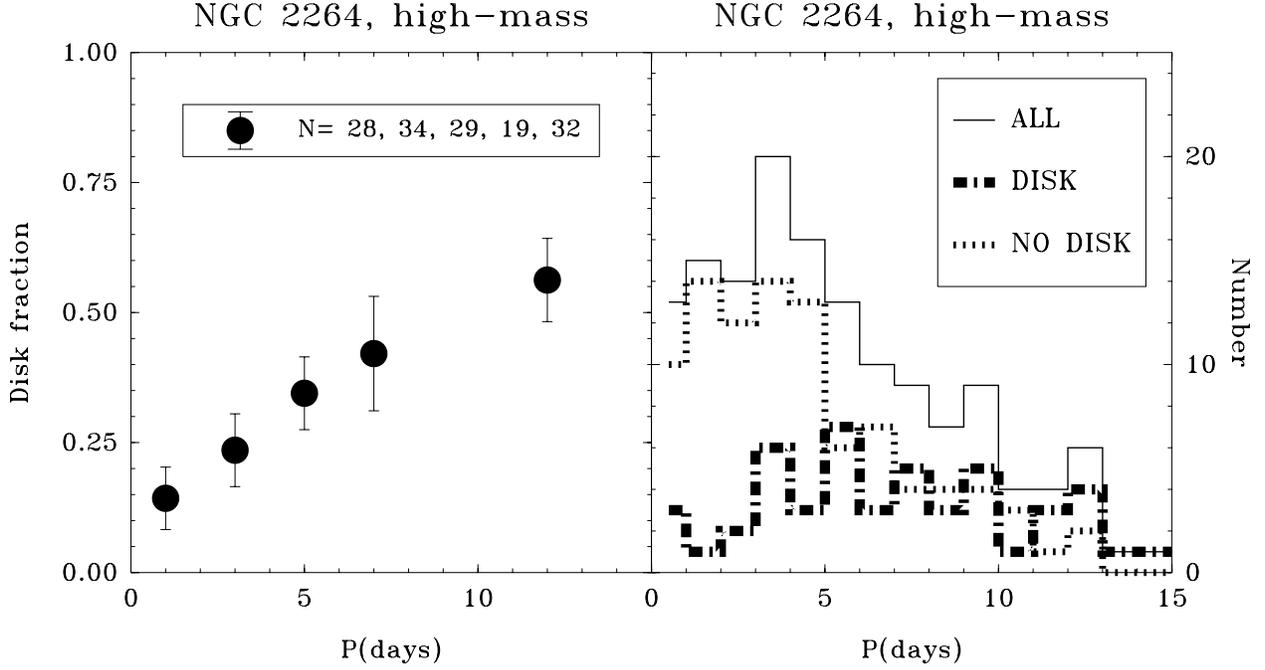}
\caption{{\bf Results for high-mass stars in NGC 2264.} {\bf (Left Panel)} The
disk fraction as a function of period for high-mass stars.  The error bars
represent the 68\% confidence level (1$\sigma$) of the measurements.  When
only high-mass stars are considered, the connection between the presence of a
disk and slow rotation becomes evident across the entire range of the period
distribution.  {\bf (Right Panel)} The period histogram for high-mass stars.
The three different lines represent all the stars (solid line) and stars with
and without a disk (dot-dash line and dotted line, respectively).  The period
distribution of disk-less high-mass stars peaks at short periods (P $<$ 5
days), while the periods of high-mass stars with disks are consistent with a
flat distribution.  These distributions are significantly different
(P=6.1e-05, Kolmogorov-Smirnov two sample test, n1=48, n2=94).  This result
suggests that stars without disks are free to spin up faster than stars with
disks.\label{NGChigh}}
\end{figure}

Using an R-I color $<$ 1.3 (corresponding to unextincted M2 stars and
stars with earlier spectral types \citep{kenyon95}) as the cutoff for
high-mass stars and plotting disk fraction as a function of period for
the high-mass sample in NGC 2264 (Fig.~\ref{NGChigh}, left panel), a
clear increase in the disk fraction with period is revealed across the
entire period range covered by the sample.  {\it Spitzer's} 8.0
$\micron$ IRAC band is the first source of data to provide unambiguous
disk identification, allowing the populations of stars with and
without disks to be separated and plotted individually.  A histogram
of the period distributions for high-mass stars with and without a
disk (Fig.~\ref{NGChigh}, right panel) shows that these distributions
are significantly different (P=6.1e-05, Kolmogorov-Smirnov two sample
test, n1=48, n2=94).  Although there is a relatively flat distribution
of stars with disks, there is a large peak of shorter-period stars
(1-5 days) with no disks and far fewer with long periods.  These
different distributions are a clear indication that star-disk
interaction regulates the angular momentum of stars as they contract
onto the main sequence.

\subsection{The ONC}
\begin{figure}
\epsscale{1}
\plotone{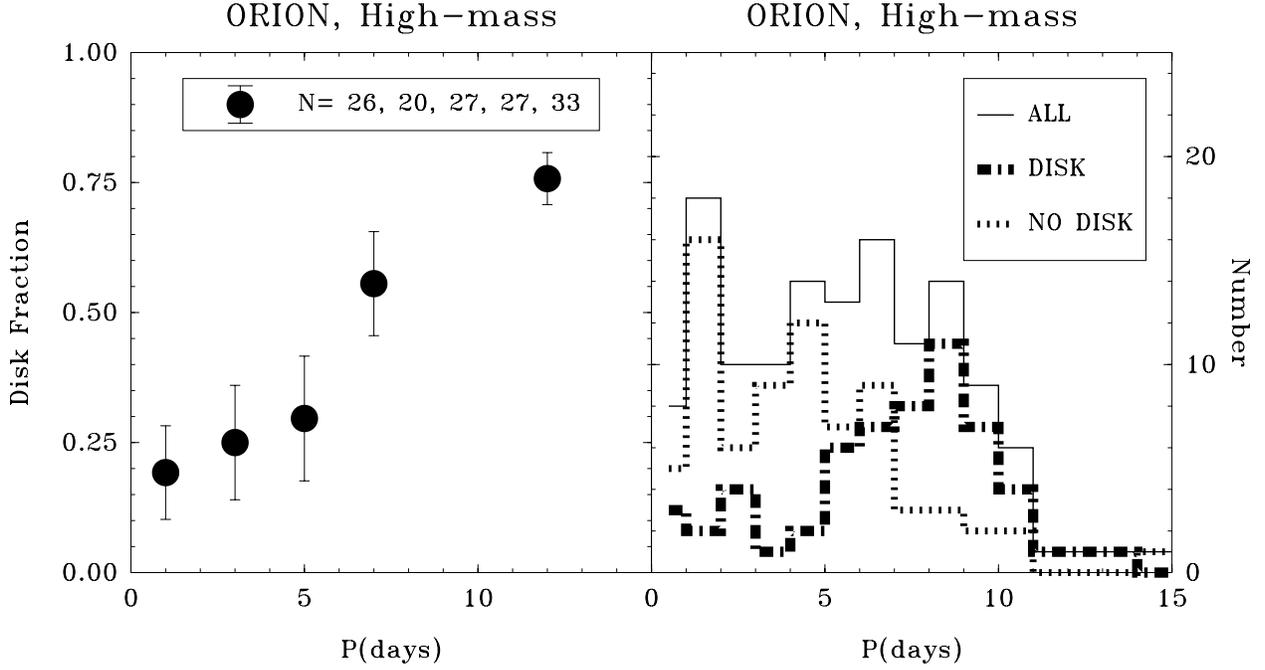}
\caption{{\bf Results for high-mass stars in Orion.} {\bf (Left Panel)} The
disk fraction as a function of period for high-mass stars with measured
spectral types in the ONC and surrounding flanking fields.  The error bars
represent the 68\% confidence level (1$\sigma$) of the measurements.  As with
NGC 2264, the disk fraction clearly increases with period across the entire
period range covered by the data.  {\bf (Right Panel)} Period histograms for
high-mass stars.  The three different lines represent all the stars (solid
line) and stars with and without a disk (dot-dash line and dotted line,
respectively).  The overall distribution is clearly a blend of the two
distinct period distributions which are significantly different from one
another (P=9.99e-07, Kolmogorov-Smirnov two sample test, n1=58, n2=75).  The
distribution of stars possessing a circumstellar disk is centered at a period
much longer than the distribution of stars with no disk.  Once again, the
result from the high-mass stars in the ONC and surrounding regions clearly
suggest that circumstellar disks are involved with angular momentum regulation
in these young stars.\label{ONChighmassall}}

\end{figure}

The evidence for angular momentum regulation through star-disk interaction is
equally dramatic in Orion if one restricts the sample studied by
\citet{rebull06} by spectral types in the literature
\citep{hillenbrand97,rebull01} to stars of spectral type M2 and earlier, even
though the sample of stars with rotation periods is cut in half as a result.
Plotting disk fraction as a function of period for the restricted sample
(Fig.~\ref{ONChighmassall}, left panel), a clear increase in the disk fraction
with period is revealed across the entire period range populated by the Orion
stars.  A histogram of the period distributions for high-mass stars with and
without a disk (Fig.~\ref{ONChighmassall}, right panel) shows that these
distributions are dramatically different (P=9.99e-07, Kolmogorov-Smirnov two
sample test, n1=58, n2=75).

\begin{figure}
\epsscale{1}
\plotone{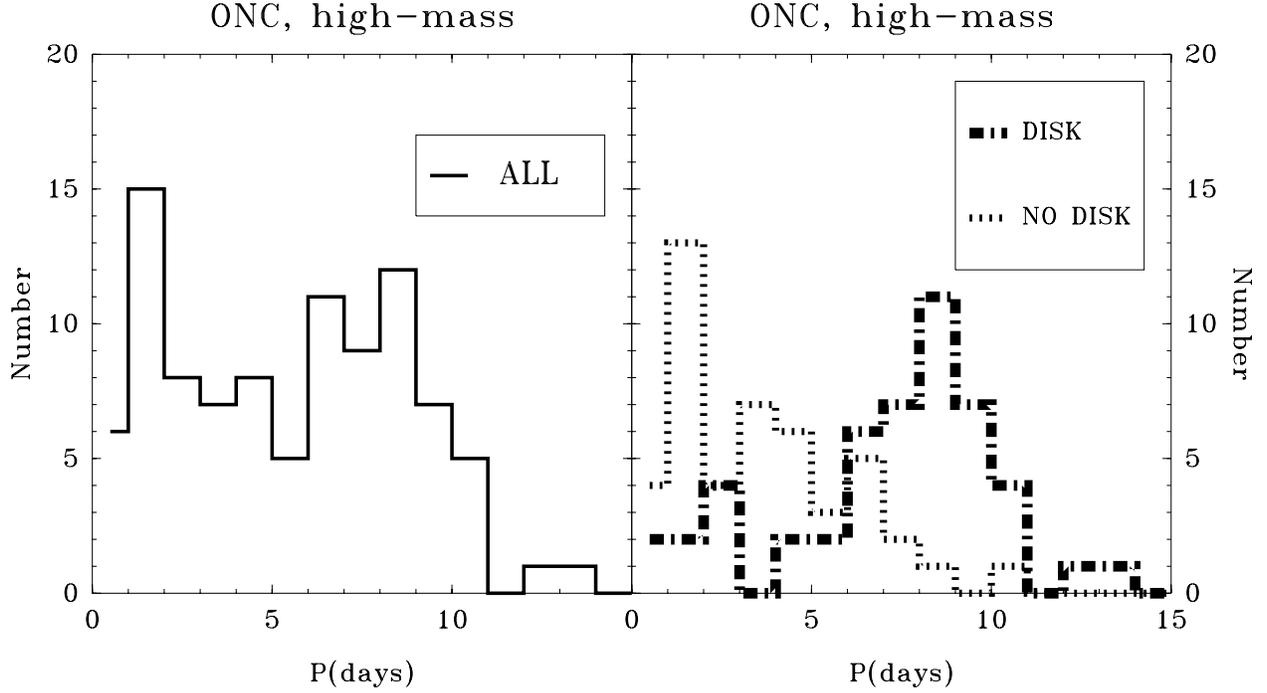}
\caption{{\bf Results for high-mass stars in the central regions of the ONC.}
{\bf (Left Panel)} The period histogram of all high-mass stars in the central
region of the ONC that have measured spectral types. {\bf (Right Panel)} The
period histograms for the same stars with (dot-dashed line) and without
(dotted line) a disk.  When restricting the sample by not including the
flanking fields, the bimodal period distribution seen by previous studies
\citep{attridge92,herbst02} is recovered.  With an accurate disk identifier
and sample selection based on spectral types, one can see that the bimodal
distribution is a blend of two dramatically different distributions, stars
with and without protoplanetary disks (P=4.3e-08, Kolmogorov-Smirnov
two-sample test, n1=49, n2=46).  The disk-less, high-mass population is
centered at a much shorter period that the population with disks, again
unambiguously supporting the picture of angular momentum regulation through
star-disk interaction.\label{ONChighmassregion}}
\end{figure}

Based on the star-disk interaction paradigm, \citet{herbst00a} predict that
the long-period peak (P $\sim$ 8 days) seen in the clear bimodal period
distribution of the high-mass stars in the ONC should be dominated by stars
with disks, while the short-period peak (P $\sim$ 2 days) should be dominated
by stars without disks.  \citet{rebull01} include stars in the ONC and in
surrounding regions termed the ``Flanking Fields,'' which are composed of
older stars than those in the younger central region of the cluster and do not
show such a clear bimodal distribution as the ONC. If one further restricts
their sample to stars in the ONC (84.1 deg $>$ RA $>$ 83.0 deg; 5.0 deg $>$
Dec $>$ -5.7 deg ) with measured spectral types (Fig.~\ref{ONChighmassregion},
left panel), one recovers the bimodal distribution seen by earlier
observations \citep{attridge92,herbst00a,herbst02}. A period histogram of
stars with disks over-plotted on a period histogram of stars without disks
(Fig.~\ref{ONChighmassregion}, right panel) reveals two distinct and cohesive
rotation period distributions, one populated by stars lacking disks peaked at
P $\sim$ 2 days and the other by stars with disks peaked at P $\sim$ 8 days,
which, blended together, form the bimodal distribution of the high-mass stars
in the ONC.  Separating and plotting individually these two populations of
stars with and without disks results in an unambiguous indication that
star-disk interaction has prevented the spin-up of PMS stars in the ONC.

\section{Discussion}\label{discussion}
\subsection{The Mass Effect}\label{meffect}

In \S\ref{results}, we have discussed the significant differences in
the period distributions of low- and high-mass stars in NGC 2264 and
the ONC.  These differences are not fully understood but can be
partially accounted for by the fact that lower mass stars of a given
age have smaller radius, R. Thus, for a given specific angular
momentum, $j$ ( $j$ $\propto$ R$^2$/P), they are in fact expected to
have a a shorter period, P \citep{herbst01}. However, since $j$ still
seems to be higher for low-mass stars than for the high-mass
counterparts, it has also been suggested that the disk regulation
mechanism is less efficient in low-mass stars than in high-mass stars
due to differences in accretion rates and the strength or structure of
their magnetic fields \citep{lamm05}.

\begin{figure}
\epsscale{1.0}
\plottwo{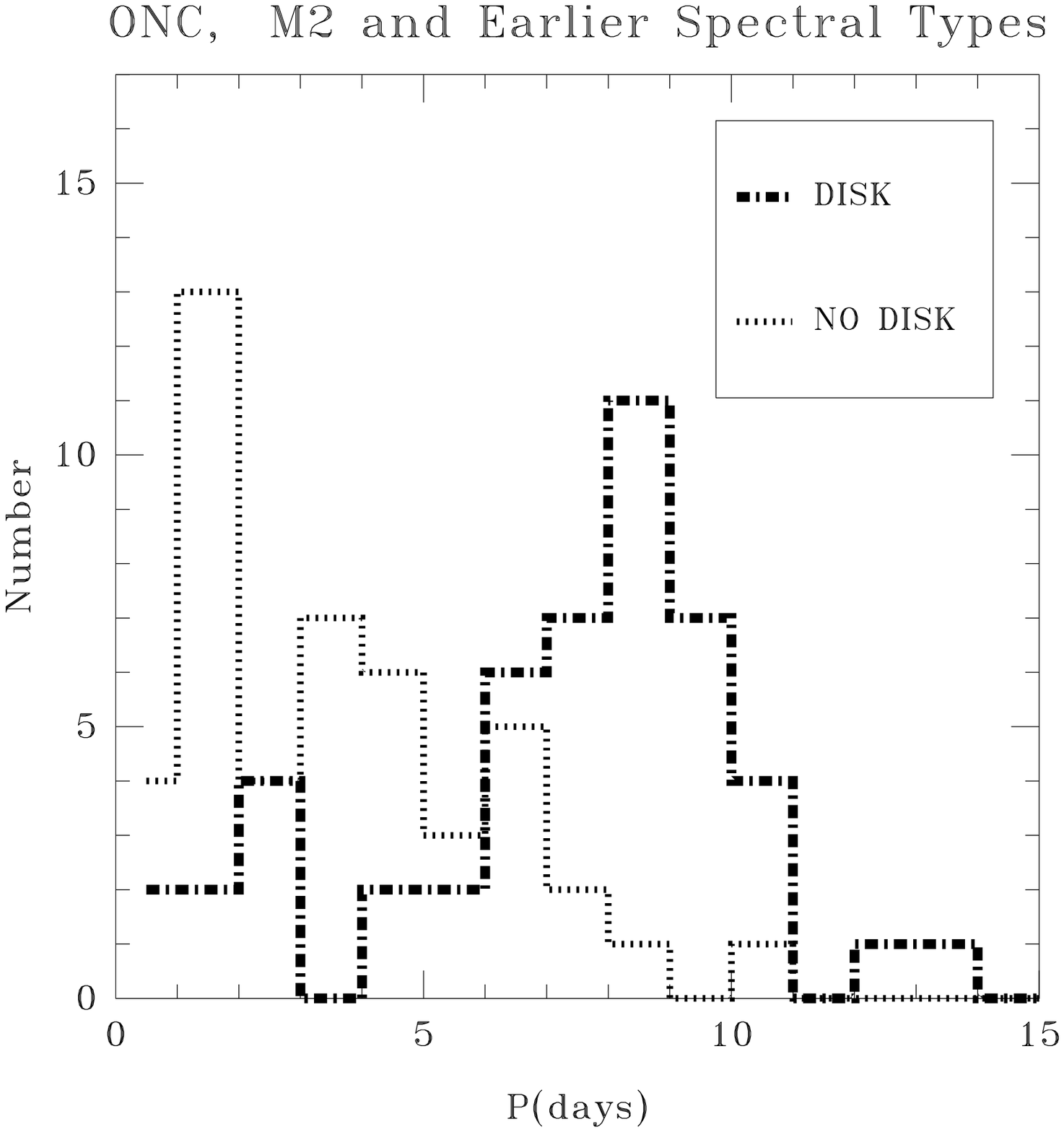}{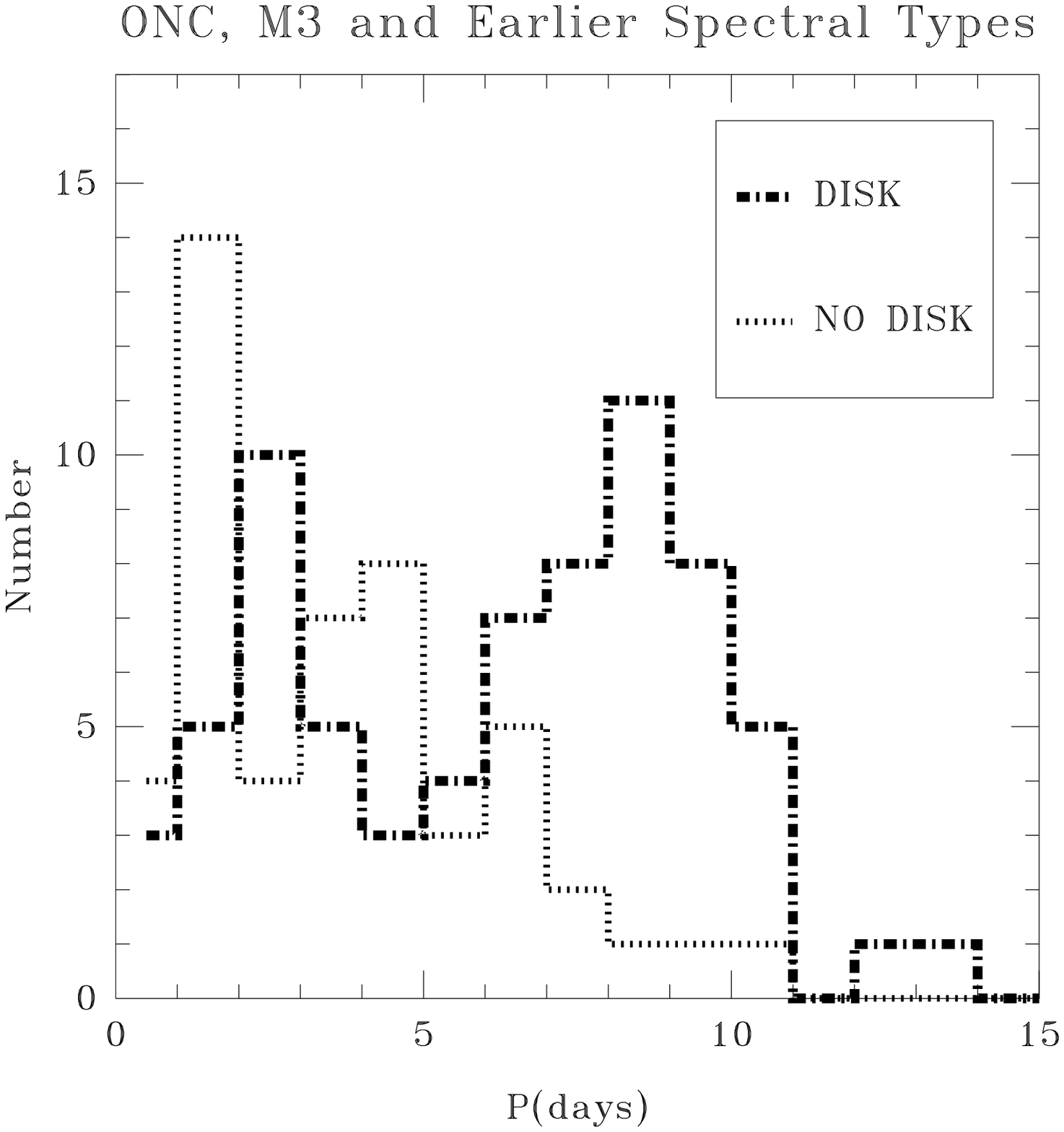}
\caption{{\bf The effect of a different mass cut on the period distribution of
    stars with and without disks in the ONC.}  {\bf (Left Panel)} Period
    histogram for high-mass stars (M2 and earlier spectral types) with (n =
    49) and without (n = 46) a disk (dot-dash and dotted line, respectively).
    {\bf (Right Panel)} The same plot with a slightly different mass cut.
    This histogram includes M3 stars (i.e. stars with slightly lower masses).
    Again, stars with disks (n=71) are represented by a dot-dash line, and
    stars without disks (n=50) by a dotted line.  This panel shows that even a
    {\it small} contamination of the high-mass star sample by stars with
    slightly lower masses will result in a short-period (P $<$ 4 days) peak of
    stars with disks that will weaken the observational signature of star-disk
    interaction on angular momentum (P increases from 4.3e-8 to 1.1e-4 in a
    Kolmogorov-Smirnov two sample test when comparing the disk and no-disk
    samples in the right panel  to those in the left.  This is due to
    the fact that low-mass stars (M3 and later spectral types) tend to have
    very short periods (P $<$ 4 days) regardless of the presence of a
    disk.\label{ONCmasscut}}
\end{figure}

It is easy to show, by making a slightly different mass cut in our ONC
analysis, that even small inaccuracies in spectral classification can lead to
a severe blending of the period distributions of stars with and without
disks. In Fig.~\ref{ONCmasscut}, we compare the mass cut we use in our
analysis (left panel) to a slightly different mass cut, including lower-mass
stars by one spectral sub-type in the sample (right panel).  Because low-mass
stars inherently rotate faster than high-mass stars, regardless of the
presence of a disk, contaminating the high-mass sample with low-mass stars
will mask the effect star-disk interaction has on PMS star rotation periods.

The extreme sensitivity of period distribution to mass is difficult to explain
in terms of slowly varying quantities such as radius and accretion rates.  The
observed dependence of rotation periods on mass is most consistent with the
picture that a \emph{sudden} change in the strength or structure of the
magnetic field at the boundary between M2 and M3 stars is responsible for the
observed differences in period distributions of low-mass and high-mass PMS
stars.  We note that, depending on the evolutionary tracks used, the masses
corresponding to given spectral types will change.  \citet{siess00} tracks,
for example, give slightly higher masses than do those of \citet{dantona98}.
However, the boundary we draw between the high- and low-mass samples is based
on spectral type, and our conclusions are not affected by the difference in
the corresponding masses derived from different evolutionary tracks.

The vast majority of stars of the masses and ages of those considered
in our study are fully convective. Unfortunately, the dynamos
operating in fully convective stars are far less understood than the
Solar-type dynamo, which operates in the boundary layer between the
convective envelope and the radiative core.  Models disagree on the
magnetic field strengths and topologies that can arise from fully
convective stars \citep{brun05,chabrier06} and are clearly not
advanced enough to predict how the strengths and topologies depend on
stellar mass \citep{donati06}. The extreme sensitivity of period
distribution to stellar mass could represent an important
observational constraint for the theoretical work in the area.

The extreme sensitivity of period distribution to mass also explains
previous {\it Spitzer} results that showed inconclusive evidence of the
star-disk interaction scenario.  \citet{rebull06} found a separate
sub-population of fast rotators with P $\le$ 2 days with a low disk fraction
(where there are few high- or low-mass stars with disks) and statistically
indistinguishable period distributions for stars with and without disks at P
$>$ 2 days (as is the case with NGC 2264 when analyzing the entire sample
instead of only high-mass stars).  The longer-period stars in those results
are a blend of high- and low-mass stars which have different period
distributions, affected by something other than star-disk interaction alone.

\begin{figure}
\epsscale{1.0}
\plotone{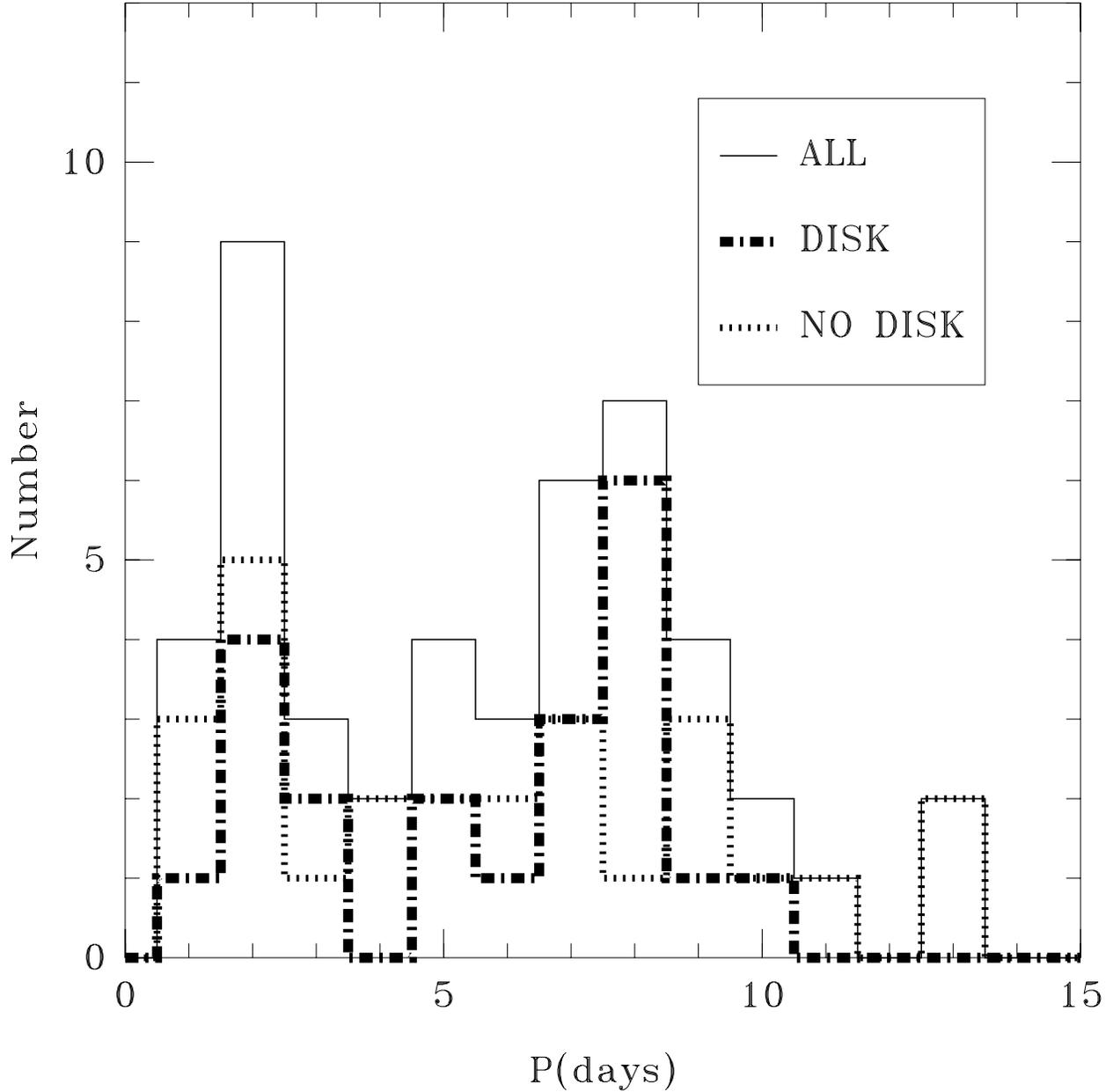}
\vskip 0.2pt
\caption{{\bf Histograms of high-mass stars in IC 348.} After dividing the IC
  348 sample of stars with known rotation periods by mass, there are too few
  stars to study the disk and no-disk populations separately.  As seen in the
  figure, very few (or no) stars remain in each period bin.  More rotation
  periods in that cluster would be needed to observe signatures of star-disk
  interaction affecting the period distributions.\label{IC348}}
\end{figure}

In paper I, we obtain the same result for IC 348 because that cluster has too
few member stars with known rotation periods to study the high- and low-mass
samples separately (Fig.~\ref{IC348}).  As a result, after isolating the small
sample, we find only a 1-$\sigma$ hint that the high-mass stars rotating
slower than the median (P = 6.2 days) have a higher disk fraction (50$\%$
$\pm$ 10$\%$ [12/24]) than the high-mass stars rotating faster than the median
(39$\%$$\pm$ 10$\%$ [9/23]). We predict that once a significant number of
rotation periods (100-150) become available for high-mass stars in IC 348 and
its surroundings, the same clear increase in disk fraction with rotation
period seen in NGC 2264 (Fig.~\ref{NGChigh}) and the ONC
(Figs.~\ref{ONChighmassall} and \ref{ONChighmassregion}) will become evident
in the IC 348 region as well.

\subsection{Outstanding Questions}

\subsubsection{Quantitative models}

Our results from \S\ref{results} show that by restricting the sample of
PMS stars studied to those with an accurately determined mass range, and by
using a reliable disk indicator like the photometry from {\it Spitzer's} IRAC
instrument, clear observational signatures of star-disk interaction become
evident. Using {\it Spitzer} mid-IR data as a disk indicator, we can finally
progress from first-order issues such as whether or not circumstellar disks
regulate the angular momentum evolution of PMS stars to ones such as what
initial conditions and PMS star and disk parameters are consistent with the
observed period distributions of stars with and without a disk, or what
constraints the observed distributions can place on disk evolution. 

For instance, comparing the observed period distributions to Monte Carlo
simulations (introduced by \citet{rebull04} and improved upon in the
discussion in Paper I) can give us information about the disk-release time of
PMS stars and the efficiency with which the disks drain their angular
momentum. Results from \citet{rebull04} suggested that a significant fraction
($\sim$30$\%$) of high-mass stars must evolve conserving angular momentum from
the time they form in order to reproduce the bimodal distribution observed in
the ONC. In the context of star-disk interaction, this implies an extremely
short disk lifetime ($<$ 1 Myr) for a significant number of stars.
Preliminary comparisons of the period distributions of stars with and without
a disk presented herein against much more detailed Monte Carlo models
\citep{cieza06a} confirm the \citet{rebull04} result. Short disk lifetimes are
also detected independently in the results of recent {\it Spitzer} surveys
\citep{padgett06,cieza07} that find that up to
50$\%$ of the youngest WTTS (age $\lesssim$ 1 Myr) show photospheric emission
in the mid-IR (8.0-24.0 $\mu$m). Our Monte Carlo models show that the period
distribution of the stars lacking disks are very sensitive to short disk
dissipation timescales because the effects of star-disk interaction are more
important at early ages when the stars undergo very rapid contraction.

The sensitivity of current PMS star rotation period distributions to short
disk dissipation time scales allows this type of numerical analysis to put
valuable constraints on both disk dissipation and planet formation time
scales, and, hence, possibly formation mechanisms. A detailed comparison of
the observed period distributions of NGC 2264 and the ONC presented herein to
Monte Carlo models will be presented in a follow-up paper (Paper III).

\subsubsection{Cluster to Cluster Comparisons}

The high mass stars in the two clusters studied in this work, the ONC and NGC
2264, have substantially different rotation period distributions. In
particular, NGC 2264 lacks the long period peak at $\sim$ 8 days and its stars
with disks show a much flatter distribution than do those in the
ONC. \citet{lamm05} argue that NCG 2264 is twice as old as the ONC and
represents a later stage in rotational evolution.  By assuming that at the age
of the ONC, NGC 2264 had the exact period distribution as the ONC has today,
they estimate that $\sim$80$\%$ of the stars in NGC 2264 have spun up from the
time it was the age of the present-day ONC until now, while only $\sim$30\%
have remained locked to their disks.

However, the difference in rotation period distributions is also likely
constrained by initial conditions and formation environment.
Characteristics such as stellar density, cluster IMF, and overall cluster mass
might play a role in the angular momentum evolution of PMS stars. The kind of
numerical models described above can be used to test whether the period
distributions observed in the ONC will naturally evolve into the period
distributions observed in NGC 2264 or if different initial conditions and
model parameters are required to reproduced the observed period distribution
of each cluster.

The observations of NGC 2264 and the ONC studied here only represent a small
fraction of the {\it Spitzer} data capable of playing a role in disentagling the
steps in the evolution of PMS stars and their disks.  {\it Spitzer} data
currently exist for tens of young nearby clusters awaiting photometric
monitoring campaigns to obtain rotation periods.  Further studies of other
clusters of different ages (from $<$1 to $>$10 Myrs) will provide a broader
age baseline with which to study the evolution of angular momentum of PMS
stars, while the study of clusters of different sizes will establish
the importance of PMS stellar environment on this evolution.

\subsubsection{Low-mass Population}
The only current complete sample of stars with rotation periods in either of
the well-studied clusters focused on in this work is the high-mass sample, or
stars of spectral type M2 and earlier. Although these stars provide a very
clear signature that star-disk interaction is regulating the spin-up of PMS
stars as they contract onto the main sequence, the whole story is as yet
untold.  Lower mass stars, half of all stars with known rotation periods,
cannot currently be studied to see if their rotation periods are similarly
affected by their circumstellar disks as no cluster has {\it Spitzer} data
deep enough to provide an unbiased sample of low-mass stars.  It is
clear that the rotation period behavior of these stars is very different from
their high-mass counterparts, but the reason for this difference is still
unknown.  These stars might have a lower overall disk fraction than high-mass
stars, which would explain the more rapid rotation of these objects.  However,
if their were no difference in the disk fraction of these stars, then
something internal to the star itself, resulting in a different magnetic
field structure for these objects, could prevent star-disk interaction from
regulating their angular momentum in the same way it does for high-mass stars.

The only cluster currently suited to be studied in this way is NGC 2264, both
because it is a rich cluster with many member stars with known rotation
period, and because it has a low enough background brightness to allow deep
{\it Spitzer} observations to detect bare stellar photospheres for the entire
periodic sample.

\section{Conclusions}\label{conclusion}

We combined stellar rotation periods of the young cluster NGC 2264 from
the literature with {\it Spitzer} photometry in order to search for the
correlation between slow stellar rotation and mid-infrared excess predicted by
disk regulation through star-disk interaction. We also re-analyzed results
from the recent \citet{rebull06} study of the ONC using the similar criteria
to those used in the NGC 2264 analysis. These two clusters combined contain
the vast majority of all known rotation periods of PMS stars.  Thanks to the
unprecedented disk detection capabilities of {\it Spitzer}, our results
provide the strongest observational evidence to date that star-disk
interaction regulates PMS star angular momentum. Our main conclusions can be
summarized as follows:
                                    
1) When stars of all masses in NGC 2264 are considered together, the only
significant result is the lower disk fraction of objects with short periods (P
$\le$ 2 days), a range that contains only $\sim$20$\%$ of the periodic stars,
with respect to that of the rest of the sample.  This is the same result found
by \citet{rebull06} for the ONC, and by itself provides only ambiguous
support for the disk regulation paradigm. However, we show that the apparent
lack of a clear overall correlation between period and IR-excess across the
entire period range is due to the strong dependence of rotation period on
stellar mass and a sensitivity bias against low-mass stars lacking disks.
                                                        
2) When only the high-mass stars (R-I $<$  1.3) in our NGC 2264 sample
are considered, the correlation between stellar rotation and IR-excess becomes
evident across the entire period range of the sample.

3) The NGC 2264 periodic sample of low mass stars (R-I $>$ 1.3) with 8.0
$\mu$m data (used for disk identification) is highly biased against the
fastest rotators. The bias in the low-mass star sample can be explained if the
fastest rotators are the lowest-mass stars in the sample and/or preferentially
have no disk.  This bias, which masks disk regulation signatures,
does not exist in the high-mass star sample. 

4) When the periodic sample of ONC stars presented by \citet{rebull06} is
restricted to high-mass stars with reliable mass estimations, the
correlation between stellar rotation and MID-IR-excess becomes apparent
across the entire range of the period distribution in the ONC sample as well.

5) We show that the long-period peak (P$\sim$8 days) of the bimodal
distribution observed for high-mass stars in the ONC is dominated by a
population of stars with disks, while the short-period peak (P $\sim$ 2 days)
is dominated by a population of stars without a disk. This result confirms one
of the main predictions of the star-disk interaction scenario
\citep{herbst00a}.

6) We argue that a quantitative comparison between the period distribution of
stars with and without a disk to numerical models is needed to constrain disk
regulation parameters such as the angular momentum transfer efficiency,
fraction of regulated stars as a function of time, etc. We will present such a
quantitative comparison to Monte Carlo models in a follow up paper (Paper
III).

7) The current samples of periodic high-mass stars in NGC 2264 and the ONC
with reliable disk indicators (e.g. [3.6]-[8.0] colors) are fairly large and
unbiased. However, accurate mass indicators (i.e., spectral types) and deeper
{\it Spitzer} observations are still needed for an unbiased quantitative
study of the role star-disk interaction plays in the evolution of low-mass
stars.

8)  Photometric monitoring of the many other young clusters already observed
    by {\it Spitzer} will reveal the importance of age and stellar formation
    environments in the angular momentum evolution of PMS stars.

\acknowledgments 
 Support
for this work, part of the {\it Spitzer} Legacy Science Program, was
provided by NASA through contract 1224608 issued by the Jet Propulsion
Laboratory, California Institute of Technology, under NASA contract
1407.

\clearpage
\tablenum{1}
\thispagestyle{empty}
\setlength{\voffset}{10mm}
\begin{deluxetable}{cccccccccccccc}
\rotate 
\tablewidth{0pt} 
\tabletypesize{\tiny}
\tablecaption{NGC 2264 Stars with periods from the
literature and Spitzer data} 
\tablehead{\colhead{RA} & \colhead{Dec} &
\colhead{Rc} & \colhead{Ic} & \colhead{Period} & \colhead{Ref} &
\colhead{Flux@3.6} & \colhead{err$_{3.6}$} & \colhead{Flux@4.5} &
\colhead{err$_{4.5}$} & \colhead{Flux@5.8} & \colhead{err$_{5.8}$} &
\colhead{Flux@8.0} & \colhead{err$_{8.0}$} \\ 
\colhead{(J2000)} & \colhead{(J2000)} & \colhead{(mag)} & 
\colhead{(mag)} & \colhead{(days)} & \colhead{} & \colhead{(mJy)} & 
\colhead{(mJy)} & \colhead{(mJy)} & \colhead{(mJy)} & \colhead{(mJy)} & 
\colhead{(mJy)} & \colhead{(mJy)} & \colhead{(mJy)}} 
\startdata 
99.94500 & 9.68167 & 13.26 & 12.69 & 3.84 & 2 & 1.06e+01 & 1.60e-01 & 
6.29e+00 & 9.98e-02 & 4.65e+00 & 1.11e-01 & 2.56e+00 & 6.06e-02\\ 
99.95292 & 9.60983 & 15.69 & 14.49 & 4.01 & 2 & 4.62e+00 & 6.79e-02
& 3.03e+00 & 3.46e-02 & 1.88e+00 & 4.59e-02 & 1.16e+00 & 3.30e-02\\ 
99.96287 & 9.60922 & 17.16 & 15.84 & 1.36 & 1 & 2.15e+00 & 3.25e-02 & 
1.40e+00 & 2.47e-02 & 1.45e+00 & 5.61e-02 & 5.89e-01 & 3.39e-02 \\
99.96954 & 9.62714 & 19.37 & 17.49 & 2.14 & 1 & 6.69e-01 & 1.15e-02 & 
5.21e-01 & 8.07e-03 & 0.00e+00 & 0.00e+00 & 2.62e-01 & 2.25e-02 \\
99.97625 & 9.94092 & 17.55 & 15.70 & 0.58 & 2 & 2.45e+00 & 3.49e-02 & 
1.84e+00 & 2.66e-02 & 0.00e+00 & 0.00e+00 & 7.91e-01 & 3.91e-02 \\
\enddata
\tablecomments{[Reference (Ref) 1 indicates periods and optical photometry
taken from \citet{lamm05}, while Ref 2 indicates periods and optical
photometry taken from \citet{makidon04}.  The complete version of this table
is in the electronic edition of the Journal.  The printed edition contains
only a sample to illustrate its content.]}
\end{deluxetable}
\setlength{\voffset}{0mm}


\clearpage

\clearpage
\tablenum{2}
\thispagestyle{empty}
\begin{deluxetable}{cccccccccccccccccccccccc}
\rotate 
\tablewidth{0pt} 
\tabletypesize{\tiny}
\tablecaption{Orion Stars detected in 3.6 and 8.0 microns with periods from
  the literature\tablenotemark{1}}
\tablehead{\colhead{Name\tablenotemark{2}} & \multicolumn{3}{c}{RA (J2000)} & \multicolumn{3}{c}{Dec (J2000)} & \colhead{F@3.6} & \colhead{err$_{3.6}$} & \colhead{F@4.5} & \colhead{err$_{4.5}$} & \colhead{F@5.8} & \colhead{err$_{5.8}$} & \colhead{F@8.0} & \colhead{err$_{8.0}$} & \colhead{Per} & \colhead{SpT} & \colhead{SpT-Ref\tablenotemark{3}} & \colhead{Mass\tablenotemark{4}} \\
\colhead{} & \colhead{ (h} & \colhead{m} & \colhead{s) } & \colhead{ (d} & \colhead{m} & \colhead{s) } & \colhead{(mag)} & \colhead{(mag)} & \colhead{(mag)} & \colhead{(mag)} & \colhead{(mag)} & \colhead{(mag)} & \colhead{(mag)} & \colhead{(mag)} & \colhead{(d)} & \colhead{} & \colhead{} & \colhead{}}
\startdata 
R01- 678 & 05 & 33 & 36.9 & -05 & 23 & 06.2 & 11.52 & 0.006 & 11.52 & 0.008 & 11.47 & 0.027 & 11.44 & 0.119 & 7.23 & ------ &  ------ & NO \\     
R01- 680 & 05 & 33 & 37.1 & -05 & 23 & 07.0 & 11.52 & 0.006 & 11.52 & 0.008 & 11.47 & 0.027 & 11.44 & 0.119 & 7.20 & ------ &  ------ & NO \\
R01- 716 & 05 & 33 & 41.6 & -04 & 55 & 59.9 & 11.91 & 0.006 & 11.90 & 0.009 & 11.78 & 0.018 & 11.88 & 0.027 & 7.55 & M5.5 & R01 & L \\
R01- 739 & 05 & 33 & 43.3 & -06 & 05 & 23.5 & 12.09 & 0.007 & 12.16 & 0.012 & 12.04 & 0.019 & 11.97 & 0.024 & 3.99 & M3.5 & R01 & L \\
R01- 749 & 05 & 33 & 44.5 & -06 & 05 & 20.5 & 12.39 & 0.009 & 12.36 & 0.011 & 12.36 & 0.020 & 12.36 & 0.031 & 15.42 & ------ & ------ & NO \\
HBC 107 & 05 & 33 & 44.9 & -05 & 31 & 08.6 & 9.92 & 0.002 & 9.97 & 0.003 & 9.89 & 0.007 & 9.82 & 0.039 & 2.64 & ------ & ------ & NO \\
Par 1266 & 05 & 33 & 46.1 & -05 & 34 & 26.5 & 10.84 & 0.003 & 10.84 & 0.004 & 10.84 & 0.011 & 10.90 & 0.039 & 4.65 & K8 & R01 & H \\
\enddata
\tablecomments{The complete version of this table
is in the electronic edition of the Journal.  The printed edition contains
only a sample to illustrate its content.}
\tablenotetext{1}{With the exception of the last 3 columns, all data come from
\citet{rebull06}.}
\tablenotetext{2}{R01 numbers come from \citet{rebull01}, HBC numbers from
 \citet{herbig88}, Par numbers from \citet{parenago54}, CHS numbers 
from \citet{carpenter01}, H97 numbers from \citet{hillenbrand97},
HBJM numbers from \citet{herbst01}, and JW numbers come from
\citet{jones88}.}
\tablenotetext{3}{R01 Spetral types come from \citet{rebull01}, while the 
H97 spectral types come from \citet{hillenbrand97}.}
\tablenotetext{4}{Stars with M2 and earlier spectral types are considered 
high-mass stars, while stars with M2.5 and later spectral types are 
considered low-mass stars.}

\end{deluxetable}
\setlength{\voffset}{0mm}


\clearpage
\pagestyle{plaintop}
\setlength{\voffset}{0mm}

\end{document}